\documentstyle[12pt]{article}
\def\be{\begin{equation}}
\def\ee{\end{equation}}
\def\bea{\begin{eqnarray}}
\def\eea{\end{eqnarray}}

\def\l{\lambda}

\author{Hans-J\"urgen Schmidt}

\title{On the Space of  3--dimensional Homogeneous 
Riemannian  Manifolds\footnote{
1991 Mathematics Subject Classification 53 B20.
Key words and phrases: Homogeneous
 Riemannian manifolds, local classification in $d=3$. This
 paper is in  final form 
and no version of it will be submitted for publication elsewhere.}}

\date{}
\begin{document}
\maketitle

\centerline{Universit\"at Potsdam, Institut f\"ur Mathematik, Am
Neuen Palais 10} 
 \centerline{D-14469~Potsdam, Germany,  E-mail:
 hjschmi@rz.uni-potsdam.de}

\begin{abstract}
 We answer the following question: 
Let $\l$, $\mu$, $\nu$ be arbitrary real numbers. 
Does there exist a  3-dimensional  homogeneous 
 Riemannian manifold whose eigenvalues  of the
 Ricci tensor are just  $\l$, $\mu$ and $\nu$  ?
\end{abstract}

\bigskip

\medskip

It is essential to observe that inspite of the 
fact  that the curvature of a 3--manifold is 
uniquely determined by the Ricci tensor, 
nevertheless, the constancy of all its three eigenvalues 
does not imply the manifold to be locally homogeneous, 
see Ref. 1 for examples. 
 The answer to the  question posed in the abstract  does
 not alter if we replace the 
word ``homogeneous" by the notion ``locally homogeneous". 
In principle, it  could have been 
answered since decades already. 
But the answer seems not to be published until this Conference. 
 And, accidentally, the answer was 
found independently by Kowalski/Nikcevic,  
Ref. 2, on the one hand, and by  Rainer/Schmidt, Ref. 3, on the other hand.

\bigskip

Of course, both  answers are equivalent. 
But the formulations are so different, that it 
is useful to compare them here. 
In Ref. 2 one defines 
 $\rho_1 = {\rm max} \{ \l, \mu, \nu      \}$,
 $\rho_3 = {\rm min} \{ \l, \mu, \nu      \}$
and  $\rho_2 =  \l + \mu + \nu - \rho_1 - \rho_3 $ and  
formulates the conditions as inequalities between the $\rho_i$.

\bigskip

In Ref. 3 the formulation uses curvature invariants; 
 first step: one observes that the answer does
 not alter, if one multiplies $\l$, $\mu$, $\nu$ 
 with the same positive constant. [Remark: this fails 
to be the 
case if we take a negative constant]; second step: we take curvature 
invariants which do not alter by this multiplication, i.e.  we look  
for suitable homothetic invariants; third step: we formulate the answer by 
use of the homothetic invariants.

\bigskip

First step: this can be achieved by a homothetic transformation of the 
metric (i.e., by a conformal transformation with constant conformal 
factor). Then one excludes the trivial case 
$\l = \mu = \nu =0$.

\bigskip

Second step: $R= \l + \mu + \nu$,   
 $N=\l^2 + \mu^2 + \nu^2$ 
 are two curvature invariants 
 with $N > 0$, and $\hat R = R/ \sqrt N$
 is a homothetic invariant. For defining a second
 homothetic invariant we introduce the trace-free part of the Ricci 
tensor
$$
S_{ij}=  R_{ij}
- \frac{R}{3} g_{ij} 
$$
and its invariant $S = S^j_i S^k_j S^i_k$. 
The desired second homothetic 
invariant reads $\hat S= S/ \sqrt{N^3}$.
   Of course, one can also 
directly define $\hat  R$ and $\hat S$ as functions of
$\l$, $\mu$ and $\nu$, e.g., $S$ is  a cubic polynomial. 
[Remark: the inequalities $S > 0$ in Ref. 3 and 
 $\rho_2 < (\rho_1+\rho_3)/2$  in Ref. 2 are
 equivalent.]

\bigskip

Third step: The answer is ``yes'' if and only 
if one of the following conditions are fulfilled:

a) $ N =0$ [flat space]

b) $\hat R^2 =3$ and $\hat S =0$ [non-flat spaces of constant curvature]

c)  $\hat R^2 =2$
 and $9 \hat R \hat S  = -1$  [real
 line

\hspace{2cm} times non-flat plane of constant curvature]

d) $\hat R^2 =1$
 and $9 \hat R \hat S  = 2$

e) $ 18 \hat S > \hat R (9  - 5 \hat R^2)$ 
  [equivalent to $\l \mu \nu >0$]

f) $  - \sqrt 3< \hat R <- \sqrt 2$  and
$$
0 < \hat S \le \left( 3 - \hat R^2 \right)^2 \, 
 \frac{5 \hat  R^2 - 6}{- 18 \hat R^3}
$$
Remarks: 1. In Ref. 2 this last 
inequality is  more elegantly written as
$$
\rho_2 \ge 
\frac{\rho_1^2 + \rho_3^2}{\rho_1 + \rho_3} \, .
$$
2. There exists a one--parameter set 
(parameter $a$ with $0 <a <1$ 
of Bianchi-type VI manifolds having
 $\rho_1 =-a-a^2$, $\rho_2 = -1-a^2$, 
 $\rho_3 = -1-a$; they represent the case where 
in subsection f) the right inequalities are fulfilled with ``$=$".]

\bigskip

{\bf Acknowledgement.} I thank for 
valuable comments by 
T. Friedrich, O. Kowalski, K. Leichtweiss, S. Nikcevic, 
M. Rainer and L. Vanhecke.

\bigskip

{\bf  References}

\noindent
1. O. Kowalski, A classification of Riemannian 
3--manifolds  with constant principal 
Ricci curvature $\rho_1  = \rho_2 \ne \rho_3$,
 Nagoya Math. J. {\bf 132} (1993) 1 - 36.

\noindent
2. O. Kowalski, S. Nikcevic, On Ricci eigenvalues of locally 
homogeneous Riemannian 3--manifolds, preprint, 
presented by S. N. at the Conf. Diff. Geom. Brno
1995, to appear in Geom. Dedicata.

\noindent 
3.   M. Rainer, H.-J. Schmidt, Inhomogeneous  
cosmological models with 
homogeneous inner hypersurface geometry, Gen. Relat. Grav. 
 {\bf 27} (1995), 1265 - 1293 and references cited there.

\bigskip
\noindent 
{\small {  In this reprint 
we removed only obvious misprints of the original, which
was published in Proc. Conf. Brno August 28 - September 1, 1995: 
Differential  Geometry and   Applications, Eds.: J.
Janyska, I. Kolar, J. Slovak; Masaryk University, Brno Czech Republic
 1996,  pages 119-120. }}
\end{document}